\documentclass[proof]{pasj00}
\Received{$\langle$reception date$\rangle$}
\SetRunningHead{Astronomical Society of Japan}{Usage of \texttt{pasj00.cls}}

\usepackage{times}
\usepackage{lineno}
\usepackage{multirow}
\begin{document}

\title{
{\it Suzaku} Observation of X-ray Variability in Soft State LMC X-1
}

\author{S. Koyama\altaffilmark{1}, 
 S. Yamada\altaffilmark{2},
 A. Kubota\altaffilmark{3},
 M. S. Tashiro \altaffilmark{1}, 
 Y. Terada \altaffilmark{1}
 K. Makishima\altaffilmark{4}
}

\altaffiltext{1}{Department of Physics
, Saitama University, Shimo-Okubo 255, Sakura, Saitama 338-8570, Japan}
\altaffiltext{2}{Department of Physics, Tokyo Metropolitan University,Minami-Osawa 1-1, Hachioji, Tokyo, 192-0397, Japan}
\altaffiltext{3}{Department of Electronic Information Systems, Shibaura Institute of Technology, 307 Fukasaku, Minuma-ku, Saitama 337-8570, Japan}
\altaffiltext{4}{Department of Physics, The University of Tokyo, 7-3-1 Hongo, Bunkyo-ku, Tokyo 113-0033}

\email{(SK):koyama@heal.phy.saitama-u.ac.jp}


\maketitle

\begin{abstract}
 This paper reports the results of {\it Suzaku} observation of the spectral variation of the black hole binary LMC X-1 in the soft state.
The Observation was carried out in 2009 from July 21 to 24.
The obtained net count rate was $\sim$ 30 counts s$^{-1}$ in the 0.5--50 keV band with $\sim$ 10\% peak-to-peak flux variation.
The time-averaged X-ray spectrum cannot be described by a multi-color disk and single Compton component with its reflection, but requires an additional Comptonized emissions.
This double Compton component model allows a slightly larger inner radius of the multi-color disk, implying lower spin parameter. 
Significant spectral evolution was observed above 8 keV along with a flux decrease on a timescale of $\sim$10$^4$--10$^5$ s.
By spectral fitting, we show that this behavior is well explained by changes in the hard Comptonized emission component in contrast to the  maintained disk and soft Comptonized emission.

\end{abstract}

\section{Introduction}

X-ray spectra from stellar mass black hole (BH) binaries are known to transit between the high/soft state (HSS) and low/hard state (LHS), 
in accordance with the mass accretion rate.
The HSS spectrum is mainly composed of a bright thermal component that dominates the soft X-ray band accompanied by a less bight power law (PL) component that exhibits a ``hard tail'' in the $\gtrsim$10 keV band.
The thermal component is well reproduced by a multi-color disk (MCD; \cite{mitsuda1984}; \cite{makishima1986}) emission from an optically thick and geometrically thin accretion disk described by the standard disk model (\cite{shakura1973}).
From the fact that the inner radius of the accretion disk derived from the
MCD model is constant (e.g., \cite{ebisawa1993}; \cite{steiner2010}), 
the existence of an innermost stable circular orbit (ISCO), 
which is determined by the BH mass and spin, has been established (see also \cite{makishima2000}).

The hard-tail component seen in the HSS  is thought to be produced by Compton upscattering of the MCD photons by energetic electrons.
This implies a situation similar to the one that produces 
the dominant hard X-ray continuum in the LHS, which is understood as resulting from thermal Comptonization of the disk photons by hot thermal electron clouds
(e.g., \cite{gierlinski1997}; \cite{frontera2001}; \cite{zdziarski2004}; \cite{makishima2008}; \cite{takahashi2008}).
However, these two phenomena show considerable differences.
The hard tail in the HSS is expressed by a single PL with a photon index of $\Gamma \sim 2.0$--$2.5$, and often extends to $\sim 1$ MeV,
in contrast to the LHS continuum which is flatter ($\Gamma < 2$) with clear a cut-off at $\sim 100$ keV.
As a result, the HSS hard tail can be explained either by non-thermal Comptonization,
or by a hybrid of non-thermal and thermal Comptonization processes (\cite{zdziarski1993}; \cite{gierlinski1997}; \cite{coppi2002}).
Overall, the hard-tail phenomenon is much less well understood than the soft component.

When a BHB (particularly a transient one) evolves from the LHS into the HSS,
it often passes through another spectral state, called the very high state (VHS; e.g, \cite{miyamoto1991}).
The VHS spectrum is characterized by luminous disk emissions with a strong PL hard tail with a steep slope of $\Gamma >$ 2.4.
The latter is resolved to two hard-tail, and explained by non-thermal plus thermal combined Comptonization 
(e.g., \cite{kubota2004b} ; \cite{kobayashi2003}), of which the thermal fraction is thought to decrease as the system approaches the HSS from the VHS.
This state evolution is also thought to involve changes in the optically-thick disk,
which extends down to ISCO in the HSS (as described above) but while is likely to be truncated in the LHS (e.g., \cite{makishima2008}).
The VHS is thus understood as a transient stage
in which the innermost disk radius gradually decreases and approaches the ISCO.
(\cite{kubota2004b}; \cite{done2006}; \cite{tamura2012}).
In this way, the optically-thick disk and the Comptonizing clouds 
are suspected to be coupled to each other through the state transition.
In the present paper, we examine possible relationships between the disk and the Compton-clouds in BHBs by examining a VHS that is rather close to the HSS.
This objective requires broadband coverage from $\sim 1$ to $\sim 100$ keV, for which {\it Suzaku} is ideal.

LMC X-1 is a persistently X-ray luminous BH binary that is 
accompanied by an O type star with an estimated mass of 32 $M_{\odot}$,
where  $M_{\odot}$ is the solar mass.
The binary parameters are known with good accuracy: 
 BH mass $M_{\rm BH}=10.9~M_{\odot}$, inclination angle $i=36^\circ 4$,
and distance $D=48$ kpc (\cite{orosz2009}).
For decades, it has been found in the HSS or VHS,
with relatively high luminosity of $\sim$10\% of the Eddington value.
Previous observations have reported a variable inner disk radius and a steep hard-tail slope of $\Gamma \sim 3 $, which are both suggestive of the VHS.
This, together with its strong X-ray variations (e.g., \cite{ruhlen2011}), 
makes the object a good source for our study.
In previous studies of LMC X-1 in energy bands below 20 keV,
the hard tail was in most cases reproduced by a simple PL feature with reflection (e.g. \cite{gou2009}; \cite{ruhlen2011}).

\citet{steiner2012} showed clear evidence for an iron line in the time-averaged {\it Suzaku} spectrum, 
and a positive correlation between the iron line flux and Compton scattering fraction obtained from RXTE observations.
Although those results suggest that the Comptonized hard-tail photons
are illuminating the disk to produce the iron line photons,
that study considered neither  combined Comptonization nor studied spectral variations in the {\it Suzaku} data.
In the present paper, we hence reanalyze the same {\it Suzaku} observation data,
attempting to see whether the inner disk radius remained constant,
and  whether the hard tail exhibited the signature of combined Comptonization.

\section{Observation and Data Reduction}
\label{sec:observation}
{\it Suzaku}, which is the fifth Japanese X-ray satellite (\cite{mitsuda2007}), carries onboard the X-ray imaging spectrometer (XIS; \cite{koyama2007}) and the hard X-ray detector (HXD; \cite{takahashi2007}; \cite{kokubun2007}).
The XIS consists of four charge-coupled device cameras placed at the focus of the X-Ray Telescope (XRT; \cite{serlemitsos2007}), covering the 0.2--12 keV energy range. 
However, one of four units is not operational (SUZAKU-MEMO 2007-08\footnote{Available at http://www.astro.isas.ac.jp/suzaku/doc/suzakumemo/suzakumemo-2007-08.pdf}).
Of the three available XIS units, two (XIS-0 and XIS-3) are front-illuminated (FI) while the other (XIS-1) is back-illuminated (BI). 
The HXD consists of PIN silicon diodes (HXD-PIN; 10--70 keV) and Gd$_2$SiO$_5$Ce scintillators (GSO; 50--600 keV). 
In 2009 from July 21 UT 18:38 to July 24 21:29, {\it Suzaku} observed the source at ``XIS nominal'' pointing position. 
The XIS was operated in standard clocking mode and ``1/4 window'' option in order to attain a time resolution of 2 s.

We used data product from {\it Suzaku} pipeline processing version 2.4.12.26 with calibration version hxd20090511, xis20090402 and xrt20080709, and software version HEADAS 6.6.2.
XIS and HXD events were screened by standard criteria. 
We discarded events collected with Earth elevation below
5$^\circ$, or with the sun irradiated-Earth elevation below 20$^\circ$ (for XIS), 
or when the spacecraft was in an orbit phase within 436 s after (for XIS)
or 180 s before and 500 s after  (for HXD) the South Atlantic Anomaly ingress/egress,
or at low-cutoff rigidity regions below 6 GV.
We accepted only the XIS events with standard grades (0, 2, 3, 4 and 6) in the analysis.

XIS spectra and light curves were extracted from a rectangle region of \timeform{8'.6}$\times$\timeform{4'.5}, attained in the 1/4 window area and centered on the image peak, while the background was also obtained from a source-free region in the same 1/4 window area.
Furthermore, we excluded a circular region of \timeform{50''} around the image peak.
According to the software {\tt AEPILEUPCHECKUP} (\cite{yamada2012}), 
the remaining signals are evaluated to have a pileup fraction of $<$1\%.
The net averaged XISs count rates were $\sim$15 counts s$^-1$ in the exposure time of $\sim$ 110 ks for each sensor.
The XIS redistribution matrix files were calculated using the {\tt xisrmfgen} tool and the ancillary response files were simulated by the {\tt xissimarfgen} tool (\cite{ishisaki2007}).

The XIS1 unfolded data deviate systematically, at energies of $\gtrsim 3$ keV, from those of XIS0 and XIS3.
This is presumably due to the residual calibration uncertainties of XIS1 (\cite{ishida2011}, subsection 7.3.3 of Suzaku technical description\footnote{Available at http://www.astro.isas.ac.jp/suzaku/doc/suzaku\_td/} ).
Therefore we did not use the XIS1 data in the following analysis.

\begin{figure}[!h]
  \begin{center}
    \FigureFile(80mm,80mm){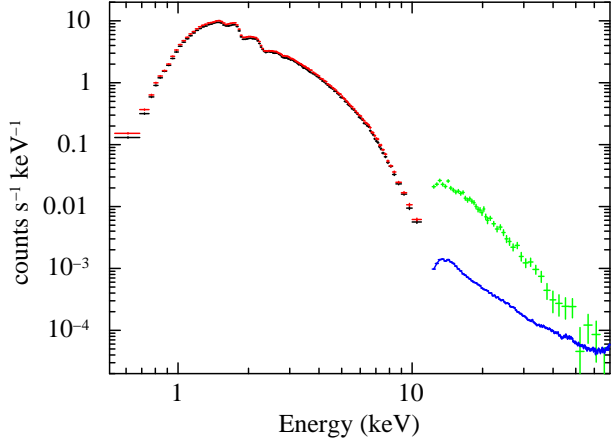}
  \end{center}
  \caption{
 Background-subtracted spectra of LMC X-1 obtained with {\it
 Suzaku} XIS0 (black), XIS3 (red) and HXD-PIN (green).
The blue line indicates the systematic uncertainty level of PIN.
 }
  \label{fig:spec_detect}
\end{figure}

The HXD screened events were used to obtain spectra and light curves from the PIN and GSO sensors. 
The net exposure of each sensor was 124 ks after dead-time correction. 
The dead-time fractions were 6.8 \% averaged for both PIN and GSO sensors.
The cosmic X-ray background was modeled assuming an exponentially cutoff PL model (\cite{boldt1987}).
Non-X-ray background (NXB) models were provided by the HXD team (\cite{fukazawa2009}). 
We used the model with METHOD=``LCFITDT(bgd\_d)'' , together with the version of  METHODV=``2.0ver0804'' and ``2.4ver0912-64'' for the data from the PIN and GSO sensors, respectively.
Figure \ref{fig:spec_detect} shows the background-subtracted spectra.
Events are significantly detected by the PIN sensor up to 50 keV, which is above the systematic uncertainty in the NXB model (3\% for PIN).

In addition to LMC X-1, there were three hard X-ray sources in the PIN field of view (FOV; \timeform{30'}$\times$\timeform{30'}); PSR B0540$-$69.3, SN 1987A and  RX J0541.4$-$6936.
Although the expected 20--100 keV flux of PSR B0540$-$69.3, $\sim 2.9\times 10^{-11} {\rm erg\ s^{-1}\ cm^{-2}}$ (\cite{campana2008}), is comparable to the observed PIN flux of $\sim 5.5\times 10^{-11} {\rm erg\ s^{-1}\ cm^{-2}}$ (modeled with a single PL), 
 it is separated from LMC X-1 by \timeform{25'}.
The PIN FOV during the present observation is evaluated to have a transmission efficiency of 21\%.
Therefore, in the following analysis, 
the spectral contribution from this active pulsar was modeled by a PL with $\Gamma$=2.12 (\cite{campana2008}), but with the flux reduced to 21\% of that reported.
In the light curve analysis, we subtracted the constant count rate of 1.21$\times$10$^{-2}$ counts s$^{-1}$ in 13--50 keV, which estimated by PIMMS\footnote{Avaliable at http://heasarc.gsfc.nasa.gov/docs/software/tools/pimms.html} tool, from the PIN light curve. 
The hard X-ray spectra of SN 1987A and RX J0541.4$-$6936 were 
cutoff-PL with $\Gamma$=1.7, cut-off energy of $E_{\rm c}$=100 keV and flux of $\sim 10^{-13}$ ${\rm erg\ s^{-1}\ cm^{-2}}$, and $\Gamma$=1.0, $E_{\rm c}$=15 keV and the flux of $\sim 10^{-12} {\rm erg\ s^{-1}\ cm^{-2}}$, respectively, as taken from the ${\it INTEGRAL}$ source catalog \footnote{Available at http://isdc.unige.ch/integral/science/catalogue} (\cite{bird2010}).
Their locations are off-axis to the PIN FOV, and the PIN efficiency was $<$50\% of the peak value.
The contributions of these two sources are thus an order of magnitude below the PIN flux and hence negligible.
Because the GSO FOV contained too many contaminating sources to individually evaluate,
 we do not use the GSO data in this paper.

\section{Analysis of Time-Averaged Spectra}

Using the X-ray spectral fitting package XSPEC version 12.8.2 
and employing the solar abundance tabulated in \citet{angr1989}, 
we analyzed 0.8--10 keV XIS spectra and 13--50 keV HXD-PIN spectra, averaged over the entire exposure.
The energy bands of 1.6--2.0 keV and 2.2--2.4 keV were excluded to avoid relatively large systematic instrumental uncertainties near the Silicon K edge (1.74 keV and 1.84 keV for K$\alpha$ and K$\beta$, respectively) and the Gold M edge (2.29 keV), respectively. 
The remaining instrumental feature at $\sim$ 3.2 keV by the Gold M edge was modeled by a Gaussian line with a fixed width of $\sigma$ = 0.1 keV (ref. \cite{kubota2007}). 
We used a combination of XIS0 and XIS3 after co-adding them together, and we added a 1\% systematic uncertainty to all XIS energy channels. 
We extended the energy range of spectral fitting to 0.1--1000 keV, as used in some convolution models. 
We fitted the XIS and PIN spectra simultaneously with a cross normalization factor of 1.16 between XIS and PIN (SUZAKU-MEMO-2008-06\footnote{Available at http://www.astro.isas.jaxa.jp/suzaku/doc/suzakumemo/suzakumemo-2008-06.pdf}).

We assumed the situation where part of the MCD photons are Compton-upscattered to the hard tail, and with reflection of Comptonized emissions.
We utilize a convolution model {\tt simpl} which takes some fraction of the disk seed photons and upscatters these to higher energy (\cite{steiner2009}).
In order to account for the reflection of the Comptonization continuum from ionized material, 
we employed the {\tt rfxconv} model (\cite{kolehmainen2011}) which is a convolution model and calculates reflection continuum and fluorescence based on the tables of \citet{ross2005}.
We fixed the ionized parameter $\xi=$1000 erg cm s$^{-1}$, inclination angle of $\theta$=36.38\degree (\cite{orosz2009}) and assumed solar abundances.
Using {\tt kdblur} (\cite{laor1991}) the reflected spectrum is smeared by general relativistic effects, with fixed inner and outer disk radii of $R_{\rm in}^{\rm kdblur} = 2\ R_g$ (\cite{gou2009}) and $R_{\rm out}^{\rm kdblur} = 400\ R_g$, inclination angle $i = 36.4\degree$, and emissivity index $\beta = -3$ ($R_{\rm g} = GM/c^2$, where $M$ is the mass of the central BH). 
The output of {\tt simpl}*{\tt diskbb} is the MCD component reduced by Comptonization and upscattered emissions.
To calculate the reflection of the Compton component,  
we divided the {\tt simpl} into {\tt simpl$_{\tt Source}$} and {\tt simpl$_{\tt Compton}$} for reduced MCD and Comptonization emission, respectively.
Consequently, we employed the continuum model as expressed by ({\tt simpl$_{\tt Source}$} + {\tt kdblur}*{\tt rfxconv}*{\tt simpl$_{\tt Compton}$})*{\tt diskbb}
modified by photoelectric absorption ({\tt phabs} in XSPEC) as Model 1.

\begin{figure*}[!h]
\begin{center}
   \FigureFile(80mm,80mm){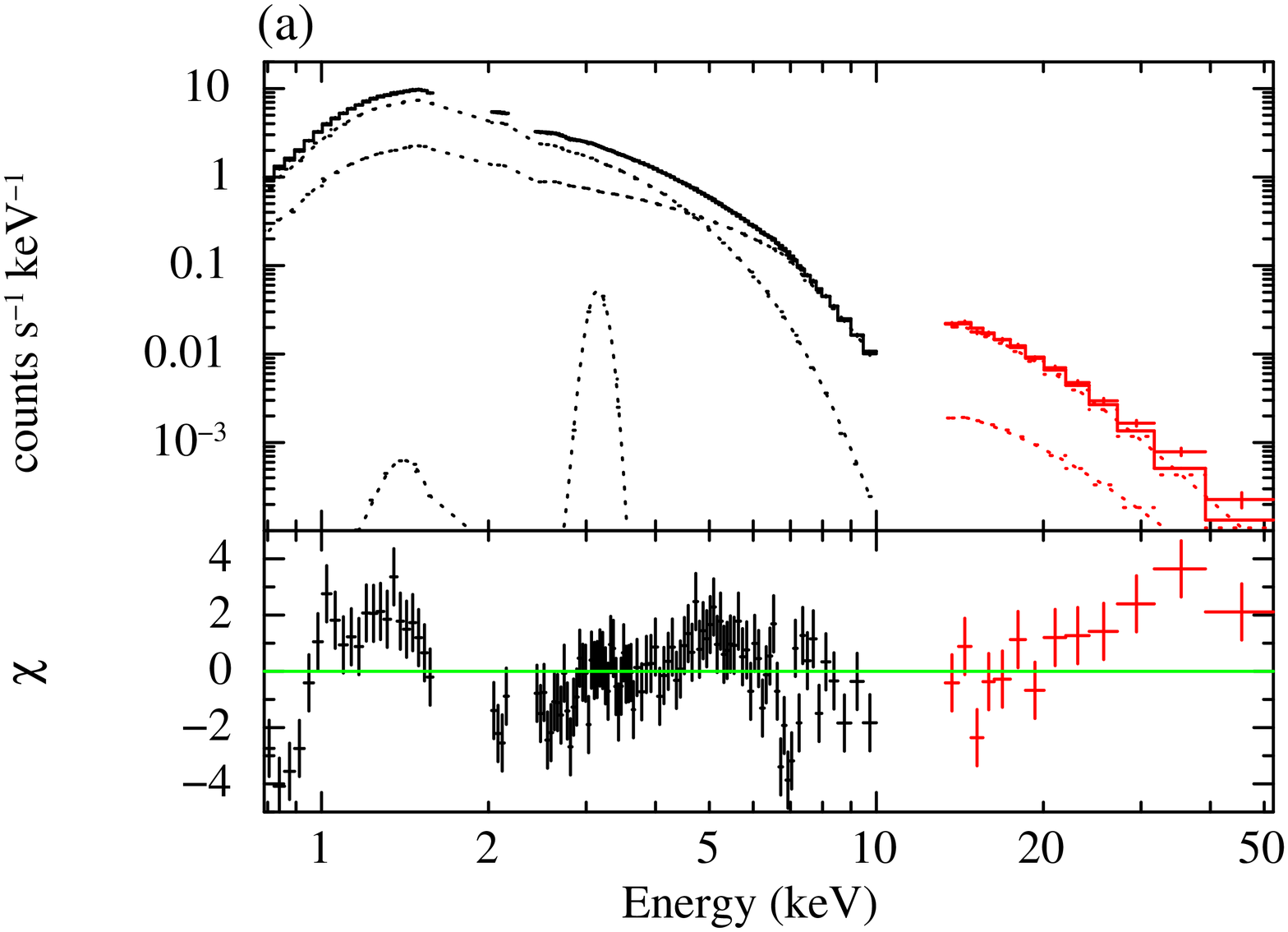}
\ \ 
   \FigureFile(80mm,80mm){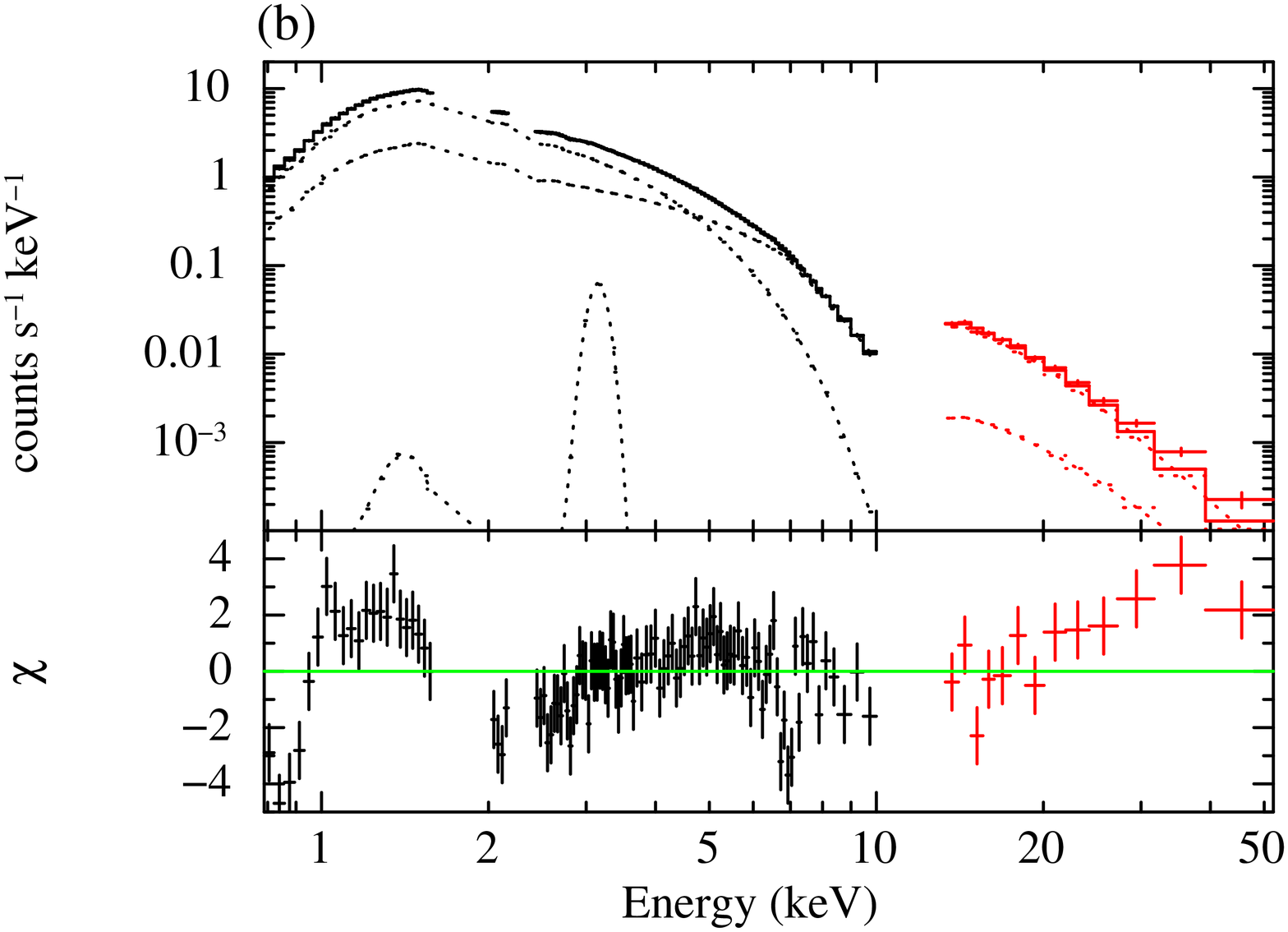}\\
   \FigureFile(80mm,80mm){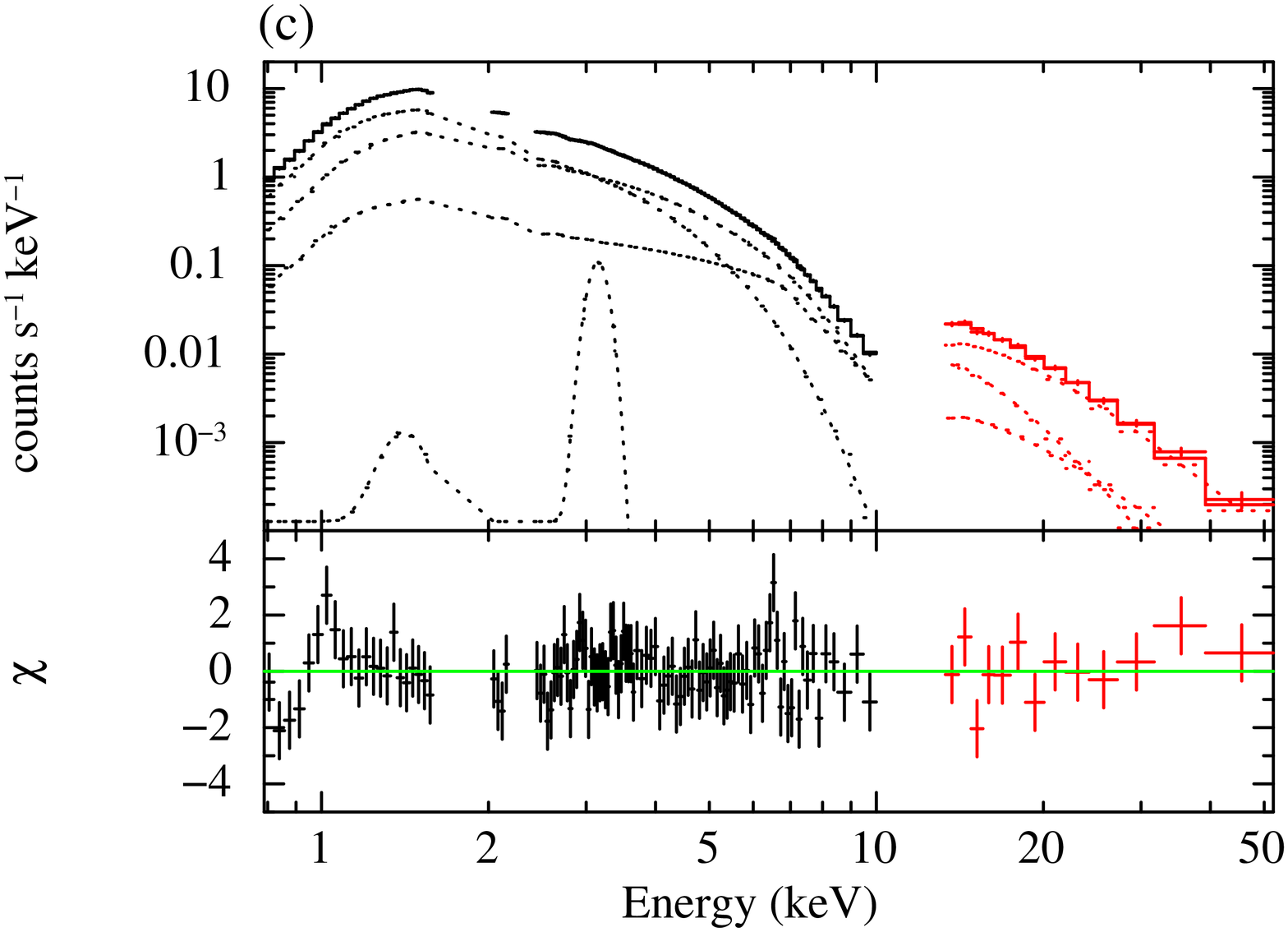}
\ \ \ \ 
 \FigureFile(80mm,80mm){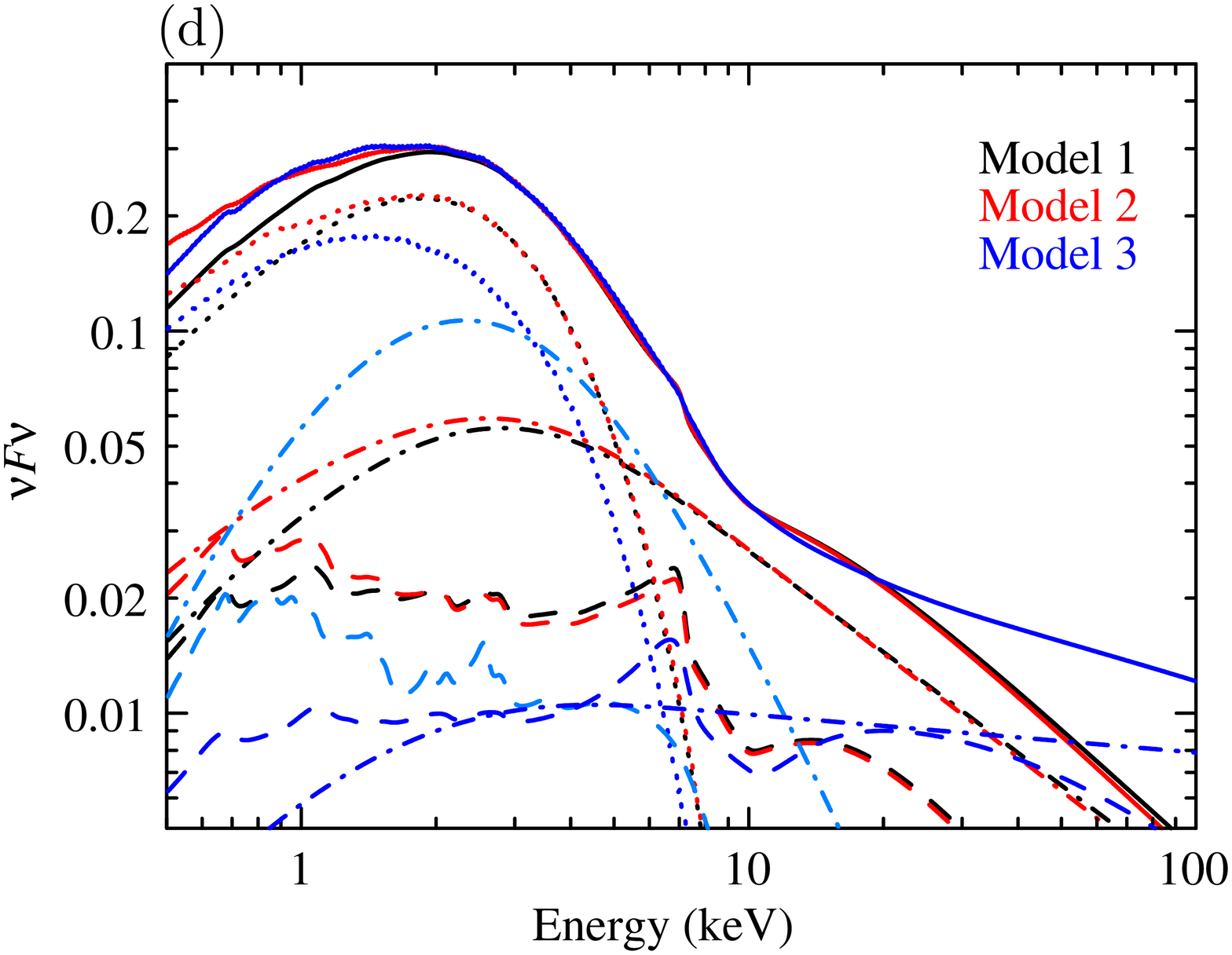}
\end{center}
 \caption{
(a) Top: Folded spectra of LMC X-1 from the best-fit for {\tt phabs} * [ ( {\tt simpl$_{\tt Source}$} + {\tt kdblur} * {\tt rfxconv} * {\tt simpl$_{\tt Compton}$} ) * {\tt diskbb} ] (Model 1).
 Black and red plots (and lines) denote the spectra (and best-fit models) for XIS(0+3) and HXD-PIN, respectively.
 Bottom: Residuals between the data and best-fit models.\\
 (b) The same as (a), but for {\tt phabs} * [ ( {\tt simpl$_{\tt Source}$} + {\tt kdblur} * {\tt rfxconv} * {\tt simpl$_{\tt Compton}$} ) * {\tt bhspec} ] (Model 2). \\
 (c) The same as (a), but for {\tt phabs} * [ {\tt simpl$_{\tt source}$} * {\tt dkbbfth$_{\tt disk}$} + {\tt kdblur} * {\tt rfxconv} * ( {\tt simpl$_{\tt HC}$} * {\tt dkbbfth$_{\tt disk}$} + {\tt dkbbfth$_{\tt SC}$} ) ] (Model 3). \\
 (d) Unabsorbed model spectra from the best-fit Model 1 (black), Model 2, (red) and Model 3 (blue).
Emission from the disk ({\tt simpl$_{\rm source}$*diskbb}, {\tt simpl$_{\rm source}$*bhspec} and {\tt simpl$_{\rm source}$*dkbbfth$_{\rm disk}$}), Compton ({\tt simpl$_{\rm Compton}$*diskbb}, {\tt simpl$_{\rm Compton}$*bhspec}, and {\tt simpl$_{\rm HC}$*dkbbfth$_{\rm disk}$}) and reflection components are represented by dotted, dash-dotted, and dashed lines, respectively.
For Model 3, the soft Compton component ({\tt dkbbfth$_{\rm SC}$}) and the reflection of that are represented by light blue.
 }
  \label{fig:spec_avr_rfxconv_dbborbhspec}
\end{figure*}

Figure \ref{fig:spec_avr_rfxconv_dbborbhspec} (a) shows the folded spectra and residuals from the best-fit model.
The best-fit parameters are shown in table \ref{tab:avr_rfxcomv_simpl_dbborbhspec} as Model 1.
This fit is unacceptable, with $\chi^2/d.o.f$ = 290.13/118, and we see large residuals below 3 keV and above 20 keV.
To check that the assumption of $\xi=1000$ erg cm s$^{-1}$ does not affect the result, we set $\xi$ to be free. 
This did not resolve the problem, with results of $\chi^2/d.o.f$ = 216.84/117, where $\xi$ = $1.43_{-0.74}^{+>98.57} \times 10^4$ erg cm s$^{-1}$ and reflection strength of $\Omega/2\pi$ = $6.07_{-0.10}^{+0.11}$, which is too large to be physically compatible.

To introduce a physical explanation for the excess in the 5--10 keV band,
we replaced the {\tt diskbb} model by {\tt bhspec} model (\cite{daivis2005}) (Model 2) to account for relativity.
Figure \ref{fig:spec_avr_rfxconv_dbborbhspec} (b) shows the folded spectra and residuals from the best-fit model.
The best-fit parameters are shown in table \ref{tab:avr_rfxcomv_simpl_dbborbhspec}  as Model 2.
The fit is still unacceptable with $\chi^2/d.o.f$ = 302.24/118.
As shown by the black and red lines in figure \ref{fig:spec_avr_rfxconv_dbborbhspec} (d), the application of relativistic effects makes the disk component slightly broader but the difference is compensated for by increased photoelectric absorption.
We tested the fit again by allowing $\xi$ to be free, and derived $\chi^2/d.o.f$ = 245.15/117, where $\xi$ = $3.67_{-2.57}^{+>96.33}$ erg cm s$^{-1}$and $\Omega/2\pi$ = $8.89_{-0.05}^{+>1.11}$.

To provide an alternative picture, we introduce a combined hard and soft Comptonization model. 
For the soft Comptonization process, we employed a {\tt dkbbfth} model (\cite{done2006}), which calculates Comptonized MCD spectrum from the accretion disk and the thermal plasma covering the inner region at $R_{\rm in} < R < R_{\rm tran}$, where $R_{\rm in}$ and $R_{\rm tran}$ are inner disk radius and outer radius of the thermal Comptonizing cloud.
The model parameters are temperature of seed MCD, electron temperature $kT_{e}$, photon index $\Gamma_{\rm th}$, $R_{\rm tran}$, and normalization.
We again include the {\tt simpl} model for hard Comptonization with fixed $\Gamma = 2.1$.
We assume that hard Comptonization occurs independently of soft Comptonization (i.e., hard electrons do not scatter soft Comptonized photons) and assume the same reflection process for both components.
This thus gives a continuum model expressed by {\tt phabs} * [{\tt simpl$_{\tt source}$} * {\tt dkbbfth$_{\tt disk}$} + {\tt kdblur} * {\tt rfxconv} * ({\tt simpl$_{\tt HC}$} * {\tt dkbbfth$_{\tt disk}$} + {\tt dkbbfth$_{\tt SC}$})] (Model 3), where the subscripts of {\tt HC} and {\tt SC} denote the hard and soft Compton component, respectively.

The fitting result with the third model is shown in figure \ref{fig:spec_avr_rfxconv_dbborbhspec} (c) and table \ref{tab:avr_rfxcomv_simpl_dbborbhspec} as Model 3.
The reduced $\chi^2$ is acceptable with 111.57/116.
Equation 1 of \citet{tamura2012} gives the relationship between $kT_e$, $\Gamma_{th}$ and optical depth assuming a source of slab geometry with seed photons at the bottom of the slab:
\begin{equation}
\tau = \frac{1}{2} \left( \sqrt{  \frac{9}{4}+\frac{3}{\Theta_{\rm e}\left( (\Gamma_{\rm th}+\frac{1}{2})^2 -\frac{9}{4} \right) }  
} 
- \frac{3}{2}
\right)
\end{equation}
where $\Theta = kT_{\rm e}/mc^2$ and the optical depth is approximately half that of spherical geometry (see also \cite{zdziarski1996}).
Using this equation, the optical depth is calculated to be $\tau$ = $0.67_{-0.13}^{+0.05}$.  
Since the normalization of {\tt dkbbfth} is not equivalent to that of {\tt diskbb},  
we calculated the inner disk radius $r_{\rm in}$ via equation A.1 in \cite{kubota2004} using the unabsorbed photon flux.
We derived $r_{\rm in} = 55.0_{-1.0}^{+1.2}$ which is consistent with the estimation from Model 1. 
This implies that the soft Compton cloud localized within $R_{\rm trans} = 8.9 R_{\rm in}$ scatters 12\% of disk photons.   

As show in figure \ref{fig:spec_avr_rfxconv_dbborbhspec} (d), the inclusion of the soft Comptonized emission shifts the disk component to lower temperature, which makes derived $r_{\rm in}$ larger substantially.
The fraction of hard Comptonized emission decreases and its photon index flatten, due to the soft Compton component. 

\clearpage

\begin{table*}[!h]
 \caption{Best-fit parameters for time averaged spectra.} 
 \label{tab:avr_rfxcomv_simpl_dbborbhspec}
\begin{center}
\begin{tabular}{llccc} \hline\hline
 &   &Model 1  &Model 2  & Model 3 \\ \hline
{\tt phabs} & $N_{\rm H}$ ($\times$10$^{21}$ cm$^2$)
     & $ 5.47 _{-0.06 }^{+0.07 }$   &  $ 6.03 _{-0.06 }^{+0.07 }$   &$ 6.06 _{ -0.11 }^{+0.09 }$    \\ \hline
{\tt diskbb}  & $kT_{\rm in} $ (keV) 
     & $ 0.784 \pm{0.006 }$  & ---   & --- \\
  & Norm  \footnotemark[$*$]
   &  $ 104.69 _{-3.49 }^{+3.94 }$  & ---  & --- \\
\hline
{\tt bhspec} & $a_{*}$     
    & --- &    $ 0.79 \pm{0.01 }$   & ---   \\
      & $L/L_{Edd}$ 
    & --- &   $0.137  \pm{0.001} $    & --- \\
\hline
{\tt dkbbfth} & $T_{\rm in}$ (keV) 
    & --- & ---  & $ 0.811 _{ -0.010 }^{+0.012 }$   \\
 & Norm ($\times 10^{-2}$) 
    & ---  & --- &  $ 1.39 _{ -0.12 }^{+0.06 }$ \\
 & $R_{\rm tran}\ (R_{\rm in})$
    & ---  & --- &  $ 8.86 _{ -0.44 }^{+0.98 }$  \\
 & $kT_{\rm e}$ (keV)
    & ---  & --- & $ 15.5 _{ -3.8 }^{+1.3 }$  \\
 & $\Gamma_{\rm th}$ 
   & --- & ---  &  $ 3.91 _{ -0.10 }^{+0.11 }$  \\
\hline
{\tt rfxconv \footnotemark[$\ddagger$]} & $\Omega/2\pi$ \footnotemark[$\S$]
    &  $ 0.64 _{-0.09 }^{+0.11 }$     & $ 0.64 _{-0.10 }^{+0.11 }$   & $ 0.98 _{ -0.18 }^{+0.34 }$    \\
\hline
{\tt simpl} & $f_{\rm scat}$ \footnotemark[$\|$] 
    &  $ 0.167 _{-0.015 }^{+0.022 }$     &  $ 0.164 _{-0.014 }^{+0.013 }$  & $ 0.033 _{ -0.007 }^{+0.002 }$     \\
 & $\Gamma$ 
   &   $ 2.88 _{-0.05 }^{+0.06 }$        &  $ 2.90 _{-0.05 }^{+0.04 }$  &  2.1 (fixed) \\
\hline
 \multicolumn{2}{l}{
 $\chi^2/d.o.f$} & 290.13/118  & 302.24/118  & 111.57/116  \\
\hline 
 Derived value & $r_{\rm in}$ (km) \footnotemark[$\#$]
    &   $54.8_{-1.0}^{+0.9}  $    &  $47.6 \pm{0.8}$  & $ 55.0 _{ -1.0 }^{ +1.2 }$ \\
  & $\tau_{\rm th}$   
     &  ---  &  ---  & $ 0.67 _{ -0.13 }^{ +0.05}$ \\
  & $f_{\rm th}$   
     &  ---  &  ---  & 0.12 \\
\hline
\\[-6pt]
\multicolumn{4}{@{}l@{}}{\hbox to 0pt{\parbox{160mm}{\footnotesize
 {\bf Notes.} The errors are 90\% confidence level for single parameter.
 \par\noindent
 \footnotemark[$*$] Normalization defined as $r_{\rm in}^2 \cos i /(D/10\ {\rm kpc})^2$
 \par\noindent
 \footnotemark[$\ddagger$] The solar abundances assumed, with fixed ionization parameter of $\xi = 1000$
 \par\noindent
 \footnotemark[$\S$] Reflection strength of the {\tt ireflect} model. $\Omega$ is solid angle  of the reflector. 
 \par\noindent
 \footnotemark[$\|$] Scattering fraction of the {\tt simpl} model
 \par\noindent
 \footnotemark[$\#$] Inner disk radius derived from fit results. \\
Model 1: Calculated from the normalization of diskbb.\\
Model 2: Equivalent to $a_*$ of  bhspec calculated with $M_{BH} = 10.9 M_{\odot}$. \\
Model 3: Estimated from unabsorbed photon flux via Equation (A.1) in \citet{kubota2004}.  
 \par\noindent 
}\hss}}
\end{tabular}
\end{center}
\end{table*}
\clearpage

\clearpage

\section{Light Curves and Spectral Variation}

\begin{figure}[!h]
  \begin{center}
   \FigureFile(80mm,60mm){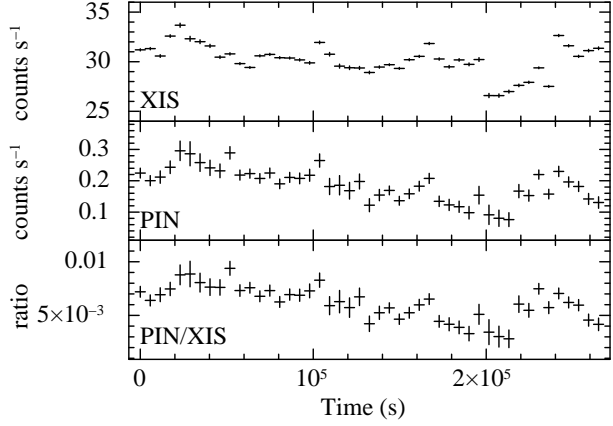}
  \end{center}

  \caption{
Background-subtracted light curves of the XIS FI in the 0.5--10 keV band and of the HXD-PIN in the 13--50 keV band, and hardness ratio variation of PIN/XIS. 
The bin size is set to a {\it Suzaku} orbital period of 5760 s.
The constant flux from PSR B0540$-$69.3 is subtracted from PIN light curve.
 }
  \label{fig:lightcurve}
\end{figure}

\begin{figure}[!h]
  \begin{center}
    \FigureFile(80mm,80mm){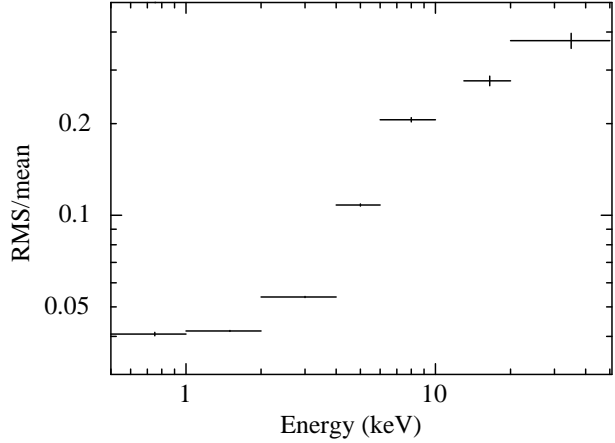}
  \end{center}
  \caption{
 Fractional rms variation spectra for XIS FI and HXD-PIN. The bin size is set to 5760 s.
 In the calculation of data points of 13--20 keV and 20--50 keV, the constant flux from PSR B0540$-$69.3 is subtracted from PIN light curve.
 }
  \label{fig:rmsspec}
\end{figure}

Figure \ref{fig:lightcurve} shows background-subtracted light curves and hardness ratio variations for LMC X-1.
Both the X-ray counts and hardness ratios exhibit variations on a time scale of $\sim 10^4 $--$ 10^5$ s.
In order to estimate the energy dependence of the variation, we calculate the rms value for each divided energy bands using a bin size of 5760 s, which is the timescale shown in figure \ref{fig:lightcurve}.
The derived rms spectrum is shown in  figure \ref{fig:rmsspec}.
The fractional rms variation clearly depends on the energy band, increasing towards higher energy range and exceeding 10\% above $\sim$6 keV.

\begin{figure*}[!h]
  \begin{center}
    \FigureFile(80mm,80mm){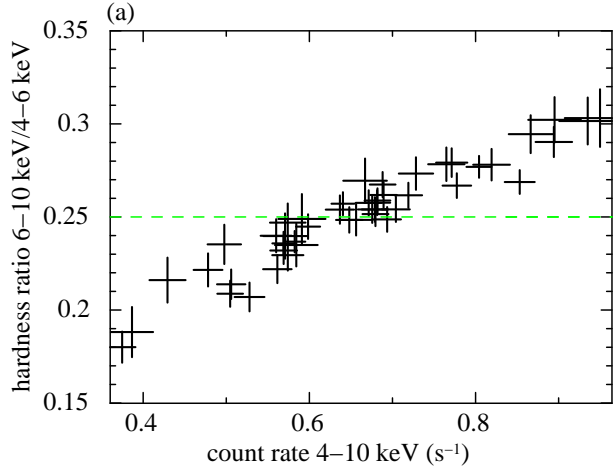}
  \end{center}
  \caption{
 Hardness ratio of 6--10 keV to 4--6 keV against 4--10 keV count rate for a bin size of 5760 s.
 The dashed line represents the average value of hardness ratio of 2.5 and indicates the limit between the hard and soft phase. \\
 }
  \label{fig:hrxcnt}
\end{figure*}

\begin{figure*}[!h]
  \begin{center}
   \FigureFile(80mm,80mm){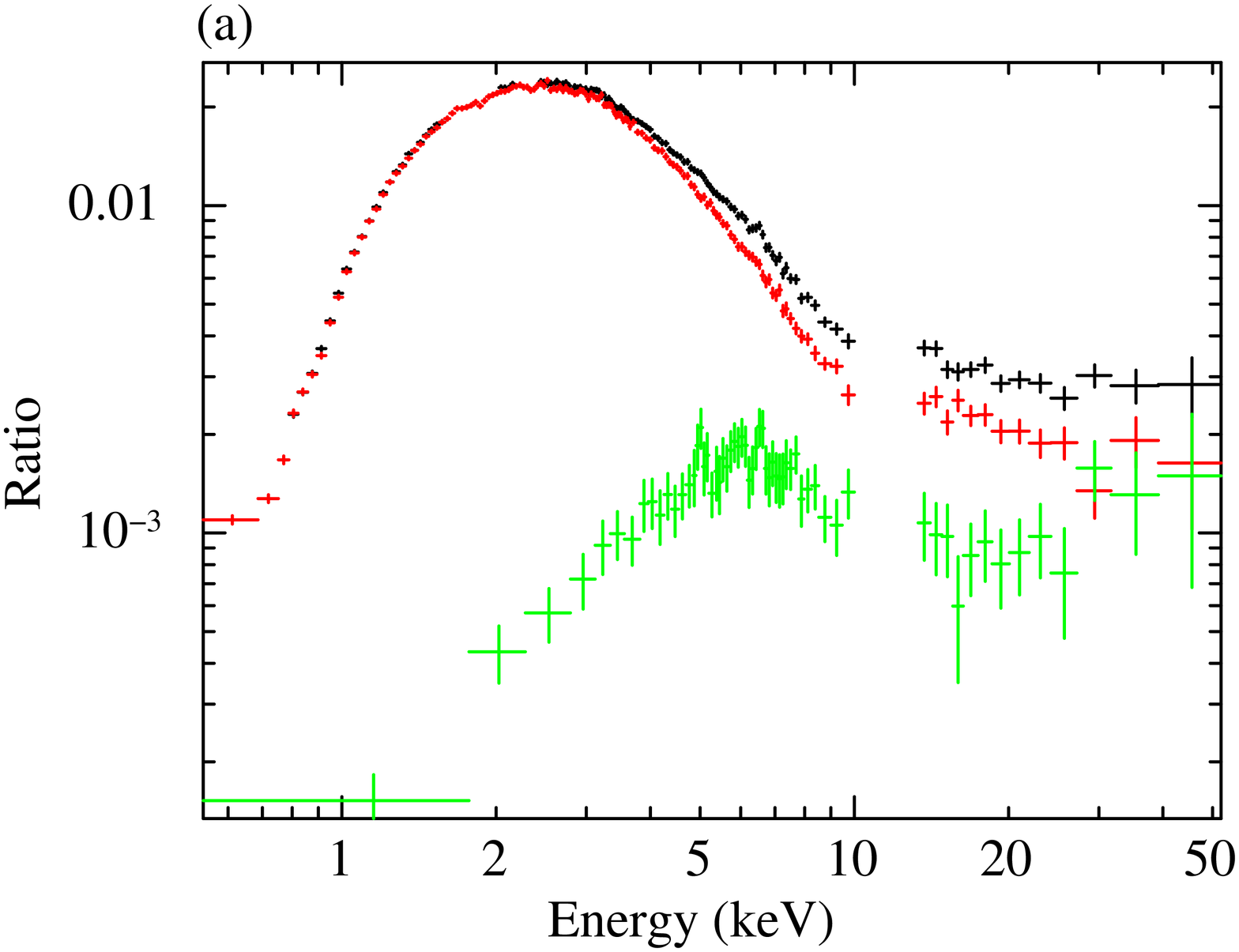}
   \FigureFile(80mm,80mm){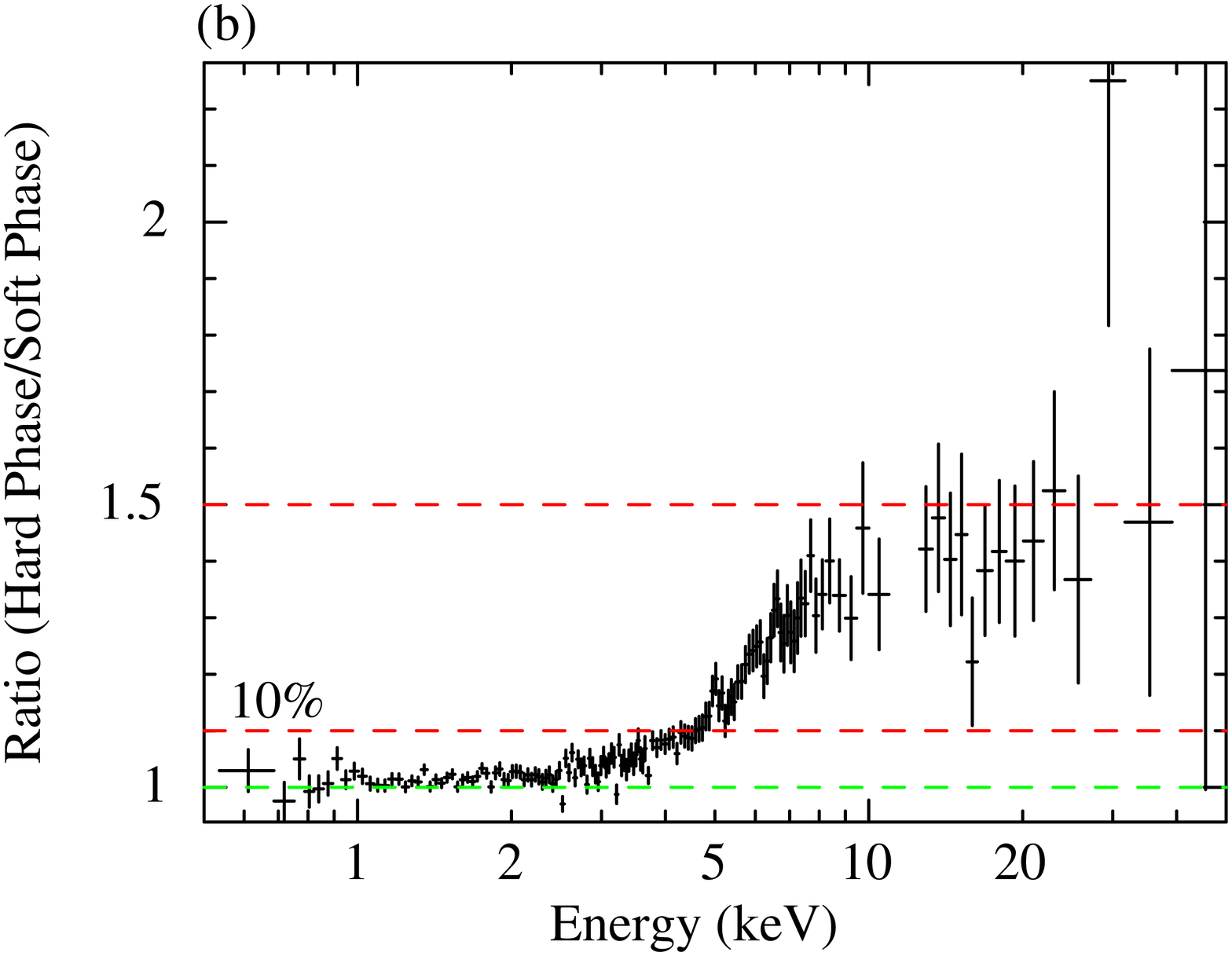}
  \end{center}
 \caption{
 (a) Spectra divided with PL ($\Gamma = 2.0$) of hard phase (black), soft phase (red) and difference between the two phases (green).
 (b) Ratio of the spectra of the hard phase to the soft phase.
 }
  \label{fig:specHSdiff}
\end{figure*}

Figure \ref{fig:hrxcnt} shows the hardness ratio of 6--10 keV to 4--6 keV against 4--10 keV count rate with a time bin size of 5760 s.
The hardness ratio is clearly correlated with source flux, which indicates significant spectral variation.
In order to evaluate the spectral variation shown in figure \ref{fig:lightcurve} and \ref{fig:rmsspec}, 
we divided the data into hard phase (HP) and soft phase (SP) for hardness ratios higher and lower than the average value of 0.25 (dashed line in figure \ref{fig:hrxcnt}).
Figure \ref{fig:specHSdiff} shows spectra normalized by PL with $\Gamma = 2$ for each phase and pulse-height ratio of HP/SP.
The differential spectrum has a peak at around 6 keV and dominates above$\sim$10 keV. 
As the pulse-height ratio shows, although the variation is $< 10\%$ in the band below 4 keV, it is $\sim$ 30--50$\%$ in the band above $\sim$8 keV.
In the time-averaged spectra (figure \ref{fig:spec_avr_rfxconv_dbborbhspec} ), the contribution of Compton and reflection components dominates above $\sim$6 keV in the energy spectrum with a peak around $\sim$4--6 keV.
The behavior thus implies spectral variability of Compton and/or reflection components.

To quantify the variations in the continuum spectra, we performed a broadband spectral fitting to each phase using Model 3.
The best-fit parameters are shown in table \ref{tab:avr_rfxcomv_simpl_dbborbhspec}.
The fit for both HP and SP are acceptable, with $\chi^2/d.o.f$ = 109.95/116 and 108.64/116 for HP and SP, respectively.
The hard Compton scattering fraction $f_{\rm scat}$ differs significantly by $\sim$40\% from $0.048_{-0.007}^{+0.008}$(HP) to $0.027_{-0.009}^{+0.008}$(SP).
In contrast, the MCD component and photo-electric absorption were stable between two phases. 
This suggests that the hard Compton cloud changes independently from the disk condition.
Although the soft Comptonization fraction $f_{\rm th}$ decreases slightly from 0.12 (HP) to 0.11 (SP),
 the changes in the other parameters $R_{\rm tran}$, $kT_e$, $\Gamma_{\rm th}$ and $\tau_{\rm th}$ are insignificant.
\clearpage

\begin{figure*}[!h]
  \begin{center}
    \FigureFile(80mm,80mm){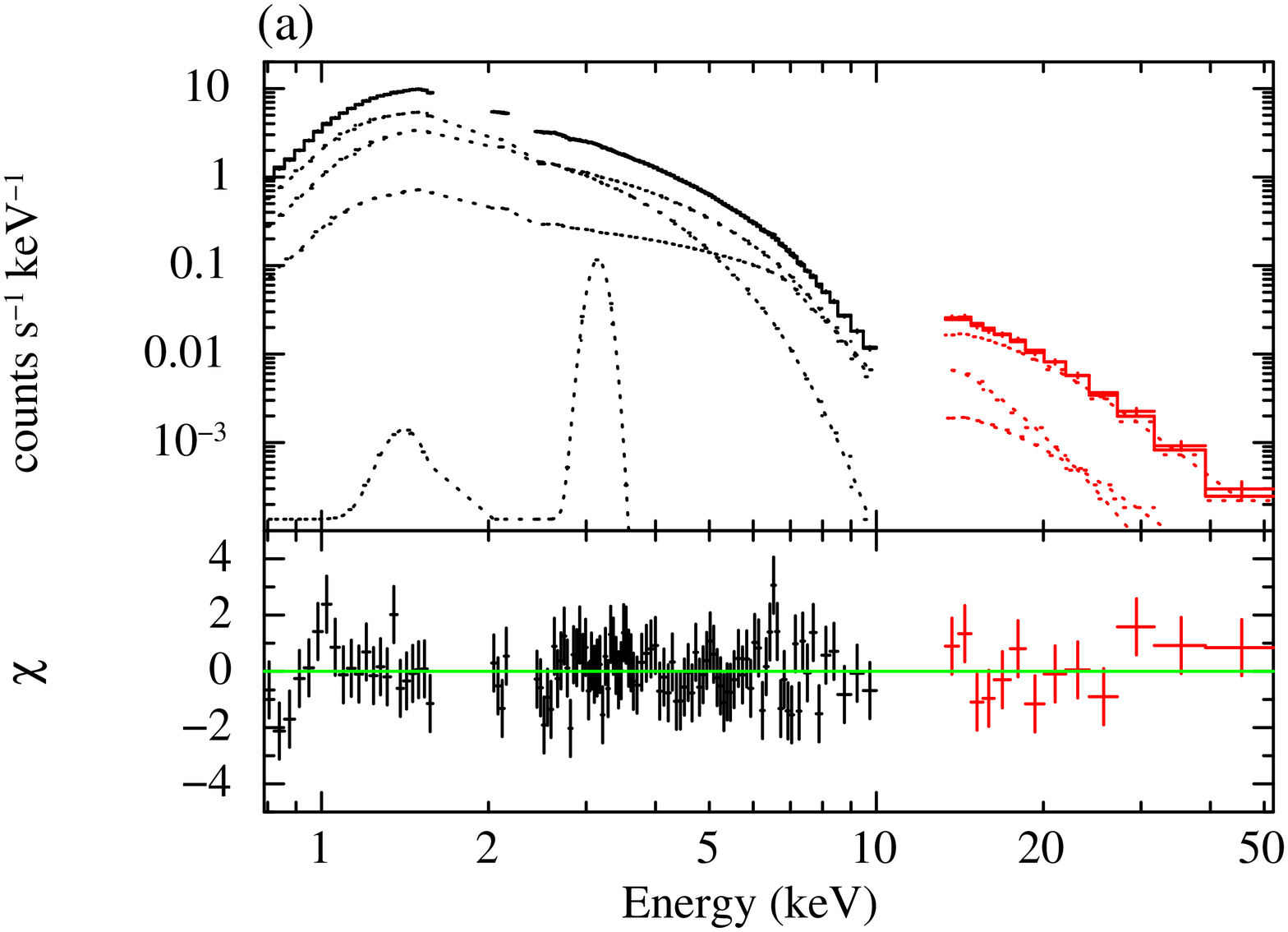} \ \ 
    \FigureFile(80mm,80mm){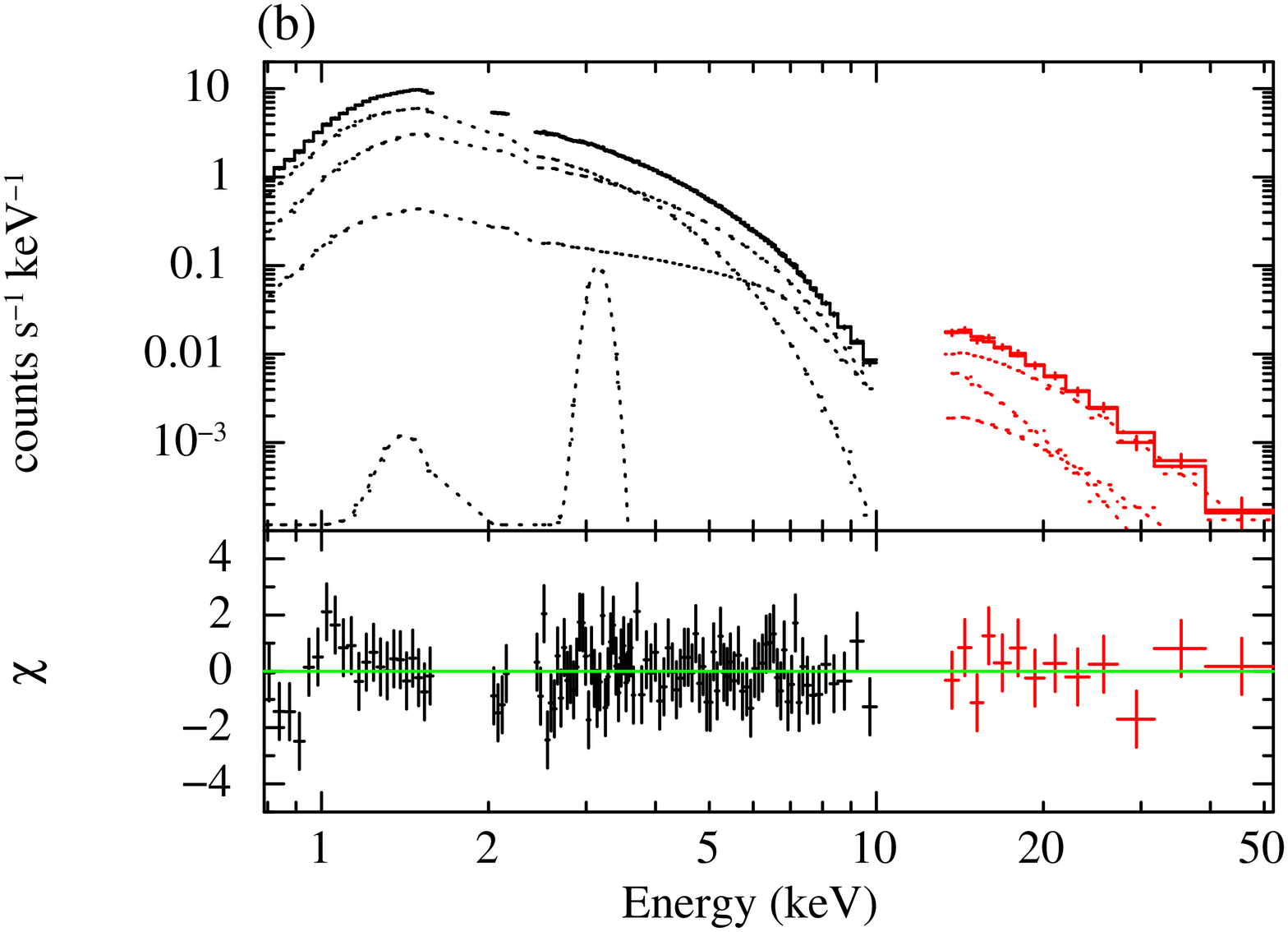} \ \ 
    \FigureFile(80mm,80mm){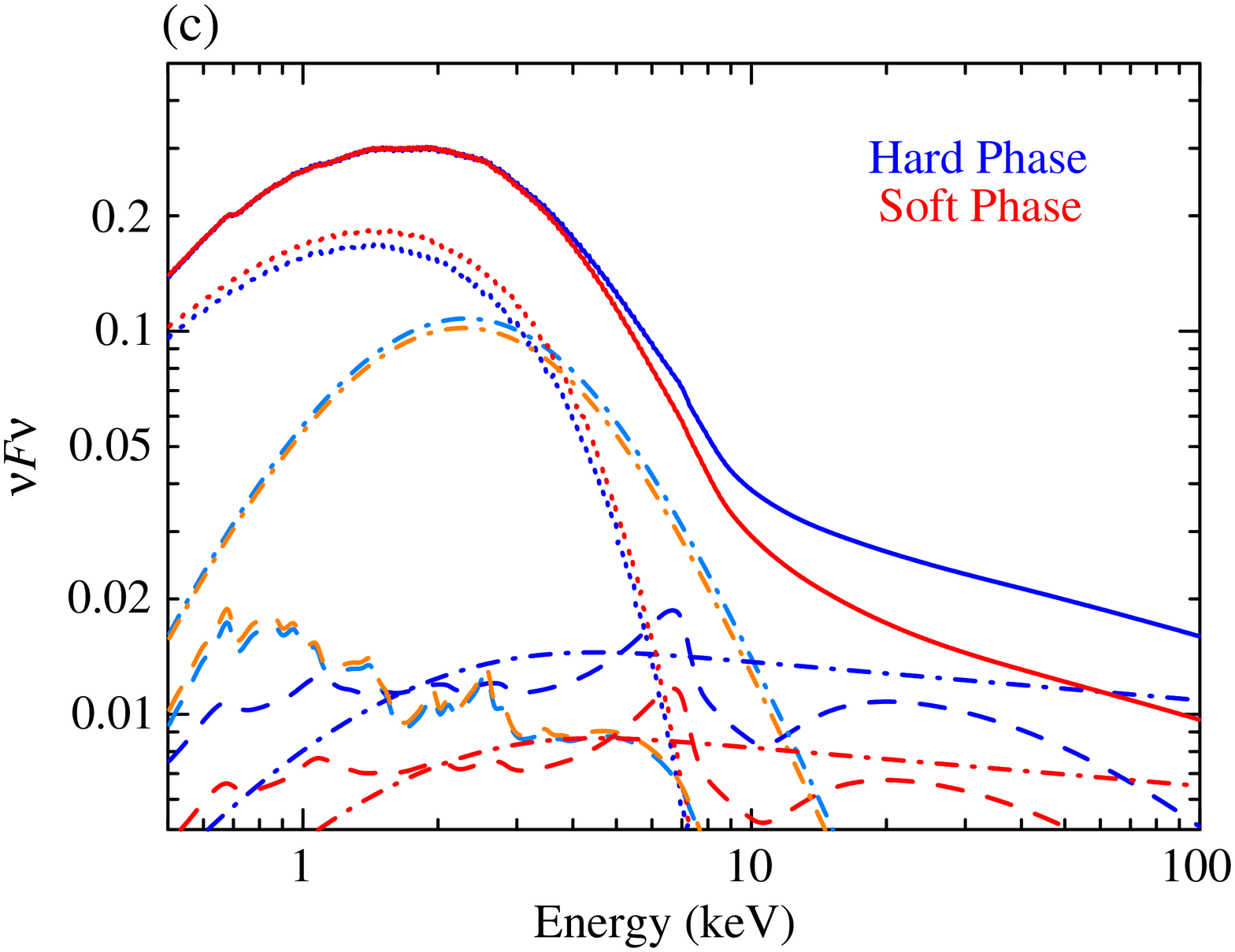}
  \end{center}
 \caption{ Unfolded spectra from the  best-fit model (Model 3) for the (a) hard phase and (b) soft phase.\\
(c) Unabsorbed model spectra of HP (blue) and SP (red) from the best fit to Model 3. 
Emission from the disk ({\tt simpl$_{\rm source}$*dkbbfth$_{\rm disk}$}), Compton ({\tt simpl$_{\rm Compton}$*dkbbfth$_{\rm disk}$}) and reflection components are represented by dotted, dash-dotted, and dashed lines, respectively.
The soft Compton compo-net ({\tt dkbbfth$_{\rm Compton}$}) and its reflection are represented by light blue and orange for HP and SP, respectively.
}
\label{fig:hardsoft_fitspec}
\end{figure*}

\begin{table*}[!h]
 \caption{Best-fit parameters for each phase.} 
 \label{tab:avr_rfxcomv_simpl_dbborbhspec}
\begin{center}
\begin{tabular}{ll cc} \hline\hline
 &   &HP  &SP  \\ \hline
{\tt phabs} & $N_{\rm H}$ ($\times$10$^{21}$ cm$^2$)
     & $ 6.09 _{ -0.13 }^{+0.01 }$   & $ 6.01 \pm{ 0.08 }$      \\ \hline
{\tt dkbbfth} & $kT_{\rm in}$ (keV) 
    & $ 0.813 _{ -0.012 }^{+0.011 }$  & $ 0.807 _{ -0.001 }^{+0.007 }$     \\
 & Norm ($\times 10^{-2}$) 
    & $ 1.35 _{ -0.08}^{+0.10}$  & $ 1.42 _{ -0.10 }^{+0.03 }$     \\
 & $R_{\rm tran}\ (R_{\rm in})$
    &  $ 8.89 _{ -0.80 }^{+0.60 }$ &  $ 9.07 _{ -0.92 }^{+1.02 }$      \\
 & $kT_{\rm e}$ (keV)
    &  $ 14.0 _{ -4.1 }^{+4.0 }$   &  $ 16.3 _{ -5.7 }^{+11.6 }$         \\
 & $\Gamma_{\rm th}$ 
   & $ 3.98 _{ -0.14 }^{+0.13 }$    &   $ 4.04 _{ -0.26 }^{+0.22 }$      \\
\hline
{\tt rfxconv \footnotemark[$\ddagger$]} & $\Omega/2\pi$ \footnotemark[$\S$]
    &  $ 0.85 \pm{ 0.24 }$     & $ 0.89 _{ -0.36 }^{+0.57 }$         \\
\hline
{\tt simpl} & $f_{\rm scat}$ \footnotemark[$\|$] 
    &   $ 0.048 _{ -0.007 }^{+0.008 }$     &   $ 0.027 _{ -0.009 }^{+0.008 }$        \\
 & $\Gamma$ 
   &   2.1 (fixed)        &    2.1 (fixed)   \\
\hline
 \multicolumn{2}{l}{
 $\chi^2/d.o.f$} &   109.95/116  & 108.64/116   \\
\hline

 Derived value & $r_{\rm in}$ (km) \footnotemark[$\#$]
    &  $ 54.1 _{ -1.2 }^{ +1.1 }$   &  $ 55.5 _{ -0.1 }^{ +0.7 }$  \\
  & $\tau_{\rm th}$   
     &  $ 0.70 \pm{0.16}$  &   $ 0.61 _{ -0.18 }^{ +0.34}$     \\
  & $f_{\rm th}$   
     &  0.12  &  0.11      \\

\hline
\\[-6pt]
\multicolumn{4}{@{}l@{}}{\hbox to 0pt{\parbox{160mm}{\footnotesize
 {\bf Notes.} The errors are 90\% confidence level for single parameter.
 \par\noindent
 \footnotemark[$\ddagger$] The solar abundances and ionization parameter of $\xi = 1000$ assumed.
 \par\noindent
 \footnotemark[$\S$] Reflection strength of the {\tt rfxconv} model. $\Omega$ is solid angle  of the reflector. 
 \par\noindent
 \footnotemark[$\|$] Scattering fraction of the {\tt simpl} model
 \par\noindent
 \footnotemark[$\#$] Inner disk radius estimated from unabsorbed photon flux via Equitation (A.1) in \citet{kubota2004}.  
 \par\noindent 

}\hss}}
\end{tabular}
\end{center}
\end{table*}

\clearpage

\section{Summary and Discussion}

We analyzed {\it Suzaku} data of LMC X-1 observed in 2009 from July 21 to 24 and obtained 0.8--50 keV spectra and light curves.
The time-averaged spectra cannot be explained by simple modeling of a non-relativistic or relativistic disk and single Comptonized emission (Models 1 and 2 in  table \ref{tab:avr_rfxcomv_simpl_dbborbhspec}) 
and the spectra exhibit large deviations below 3 keV, around 6 keV, and above 20 keV.
These residuals are successfully compensated for by adding a hard and soft Comptonization model (Model 3). 
The hard and soft components are described by a PL tail with $\Gamma$ = 2.1 representing the excess above 20 keV from the  {\tt simpl} model and 
thermal Comptonization at inner disk region of $\lesssim 9\ R_{\rm in}$ with $kT_{\rm e}\ \sim$  16 keV and $\tau_{\rm th}\ \sim$ 0.7 to describe the 4--10 keV residuals by the {\tt dkbbfth} model (figure \ref{fig:spec_avr_rfxconv_dbborbhspec}), respectively.
Using the best-fit parameters, the unabsorbed 0.8--50 keV flux is estimated to be 8.7$\times$10$^{-10}$ ergs cm$^{-2}$ s$^{-1}$, which gives an X-ray luminosity of 2.4$\times$10$^{38}$ ergs s$^{-1}$ for an isotropic emission from 48 kpc.
The bolometric luminosity was also calculated to be  3.3 $\times$10$^{38}$ ergs s$^{-1}$.

In contrast to the case where the spectrum is not fitted by the single Compton model,
\citet{steiner2012} reported reduced $\chi^2\sim$1 by similar fitting to Model 2 from the time-averaged {\it Suzaku} spectrum of LMC X-1.
This difference could be mainly caused by 
absorption model for inter-stellar medium (ISM).
They employed the absorption model of {\tt TBvarabs} (\cite{wilms2000}) with ISM composition taken from \citet{hanke2010}, while 
we employed {\tt phabs} in the analysis.
Replacing the {\tt phabs} to {\tt TBvarabs}, we obtain improvement of reduced $\chi^2_{\nu}$ from 2.6 to 1.5.
However, Model 3 still gives the best fit results with the reduced $\chi^2_{\nu}$ of 1.1.
Therefore these difference of absorption models dose not affect to the conclusion of this paper.

We found significant variation both in the XIS and PIN band. 
The count rates correlate with the  hardness ratio, and 
different spectral changes appeared in the three energy bands of below 2 keV, above 10 keV, and 2--10 keV, which exhibited variations of $\sim$1\% , $\sim$40\%, and values in between, respectively (figure \ref{fig:specHSdiff}b). 
This implies the existence of three spectral components, which is consistent with the modeling of MCD and combined hard and soft Comptonized emission. 
From the spectral fittings, we found that 
the spectral change is 
mainly explained by variations in the hard Comptonization fraction by a factor of $\sim40\%$ in contrast to keeping the MCD parameters of $T_{\rm in}$ and $R_{\rm in}$.

The obtained results show that the spectrum is composed by three spectral components of disk and two Comptonized emissions instead of disk with single Comptonized emission.
This increases inner disk radius from 47.6$\pm$0.8 km (Model 2) to 55.0$_{-1.0}^{+1.2}$ km (Model 3), hence, implies lower BH spin.

\begin{figure}[!h]
  \begin{center}
   \FigureFile(80mm,80mm){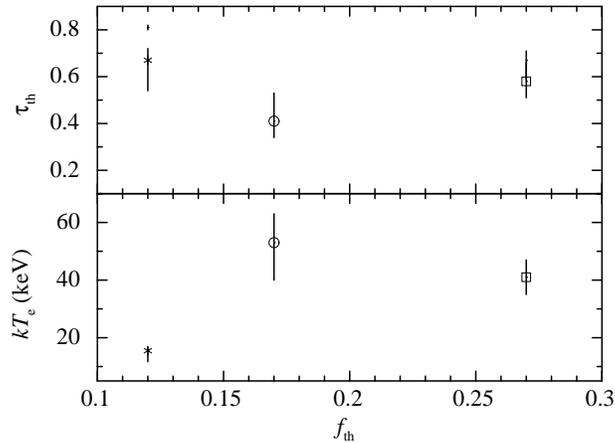}
  \end{center}
 \caption{
 Comparison of derived {\tt dkbbfth} parameters of $\tau_{\rm th}$, $kT_{\rm e}$ and $f_{\rm th}$ from three objects.
 Squares, circles and asterisks represent the parameters of GX 399$-$4(\citet{tamura2012}), 4U1630$-$47 (\citet{hori2014}) and LMC X-1 (this work), respectively.
 }
  \label{fig:comp_par}
\end{figure}

Figure \ref{fig:comp_par} shows a comparison of 
the best-fit parameters of soft Comptonized emission 
from three VHS BHBs observed by {\it Suzaku}, GX 399$-$4 (\cite{tamura2012}), 4U1630$-$47 (\citet{hori2014}), and LMC X-1 (this work).
The parameters of $kT_{\rm e}$ and $f_{\rm th}$ derived from LMC X-1 are lower than those of the other studies, although $\tau_{\rm th}$ is similar.
We also found a stable $r_{\rm in}$ at $\sim$55 km (corresponds to $\sim$3--4 $R_{\rm g}$ with $M$ = 10.9 $\MO$), which is assumed to reach the ISCO, of LMC X-1. 
In contrast, the truncated inner disk radii of GX339$-$4 and 4U1630$-$47 were reported by \citet{tamura2012} and \citet{hori2014}, respectively.
Taking into account that the inner disk radius and the soft Comptonized emission decrease through the state evolution from VHS to HSS, 
we found the feature that
 LMC X-1 is in the VHS rather than HSS, 
 and the inner disk radius reaches the ISCO though the relatively weak soft Comptonized emission.

\
\\

This work is supported by a Grant-in-Aid for JSPS Fellows (No. 258352).


\begin{thebibliography}{}


\bibitem[Anders \& Grevesse (1989)]{angr1989} Anders, E., \& Grevesse, N. 1989, Geochimica et Cosmochimica Acta, 53, 197

\bibitem[Bird et al.(2010)]{bird2010} Bird, A.~J., Bazzano, A., Bassani, L., et al.\ 2010, \apjs, 186, 1 

\bibitem[Boldt et al. (1987)]{boldt1987} Boldt, E., et al., 1987, \physrep, 146, 215

\bibitem[Campana et al. (2008)]{campana2008} Campana, R., et al., 2008, \mnras, 389, 691

\bibitem[Coppi\& Maccarone(2002)]{coppi2002} Coppi, P., \& Maccarone, T.\ 2002, APS Meeting Abstracts, 17079 

\bibitem[Davis et al. (2005)]{daivis2005} Davis, S. W., Blaes, O. M., Hubeny, I, \& Turner, N. J. 2005, \apj, 621, 372

\bibitem[Done \& Kubota (2006)]{done2006} Done, C., \& Kubota, A. 2006, \mnras, 371, 1216

\bibitem[Ebisawa et al.(1993)]{ebisawa1993} Ebisawa, K., Makino, F., Mitsuda, K., et al.\ 1993, \apj, 403, 684 

\bibitem[Frontera et al.(2001)]{frontera2001} Frontera, F., Zdziarski, A.~A., Amati, L., et al.\ 2001, \apj, 561, 1006 

\bibitem[Fukazawa et al. (2009)]{fukazawa2009} Fukazawa, Y., et al., 2009, \pasj, 61, S17

\bibitem[Gierlinski et al.(1997)]{gierlinski1997} Gierlinski, M., Zdziarski, A.~A., Done, C., et al.\ 1997, \mnras, 288, 958 

\bibitem[Gierli{\'n}ski et al.(1999)]{gierlinski1999} Gierli{\'n}ski, M., Zdziarski, A.~A., Poutanen, J., et al.\ 1999, \mnras, 309, 496 

\bibitem[Gou et al.(2009)]{gou2009} Gou, L., McClintock, J.~E., Liu, J., et al.\ 2009, \apj, 701, 1076 

\bibitem[Hanke et al.(2010)]{hanke2010} Hanke, M., Wilms, J., Nowak, M.~A., Barrag{\'a}n, L., \& Schulz, N.~S.\ 2010, \aap, 509, LL8 

\bibitem[Hori et al.(2014)]{hori2014} Hori, T., Ueda, Y., Shidatsu, M., et al.\ 2014, \apj, 790, 20 

\bibitem[Ishida et al.(2011)]{ishida2011} Ishida, M., Tsujimoto, M., Kohmura, T., et al.\ 2011, \pasj, 63, 657 

\bibitem[Ishisaki et al. (2007)]{ishisaki2007} Ishisaki, Y., et al., 2007, \pasj, 59, S113

\bibitem[Kobayashi et al.(2003)]{kobayashi2003} Kobayashi, Y., Kubota, A., Nakazawa, K., Takahashi, T., \& Makishima, K.\ 2003, \pasj, 55, 273 

\bibitem[Kokubun et al. (2007)]{kokubun2007} Kokubun, M., Makishima, K., Takahashi, T., et al.\ 2007, \pasj, 59, 53 

\bibitem[Kolehmainen et al. (2011)]{kolehmainen2011} Kolehmainen, M., Done, C. \& Diaz Trigo, M., 2011, \mnras, 416, 311

\bibitem[Koyama et al. (2007)]{koyama2007} Koyama, K., Hyodo, Y., Inui, T., et al.\ 2007, \pasj, 59, 245

\bibitem[Kubota \& Makishima(2004)]{kubota2004} Kubota, A., \& Makishima, K.\ 2004, \apj, 601, 428 

\bibitem[Kubota \& Done(2004)]{kubota2004b} Kubota, A., \& Done, C.\ 2004, \mnras, 353, 980 

\bibitem[Kubota et al.(2007)]{kubota2007} Kubota, A., Dotani, T., Cottam, J., et al.\ 2007, \pasj, 59, 185 

\bibitem[Laor (1991)]{laor1991} Laor, A., 1991, \apj, 376, 90

\bibitem[Makishima et al. (1986)]{makishima1986} Makishima, K., et al., 1986, \apj, 308, 635

\bibitem[Makishima et al.(2000)]{makishima2000} Makishima, K., Kubota, A., Mizuno, T., et al.\ 2000, \apj, 535, 632

\bibitem[Makishima et al.(2008)]{makishima2008} Makishima, K., Takahashi, H., Yamada, S., et al.\ 2008, \pasj, 60, 585 

\bibitem[Mitsuda et al. (2007)]{mitsuda2007} Mitsuda, K., Bautz, M., Inoue, H., et al.\ 2007, \pasj, 59, 1 

\bibitem[Mitsuda et al. (1984)]{mitsuda1984} Mitsuda, K., et al., 1984, \pasj, 36, 741

\bibitem[Miyamoto et al.(1991)]{miyamoto1991} Miyamoto, S., Kimura, K., Kitamoto, S., Dotani, T., \& Ebisawa, K.\ 1991, \apj, 383, 784 

\bibitem[Orosz et al.(2009)]{orosz2009} Orosz, J.~A., Steeghs, D., McClintock, J.~E., et al.\ 2009, \apj, 697, 573 

\bibitem[Ross \& Fabian(2005)]{ross2005} Ross, R.~R., \& Fabian, A.~C.\ 2005, \mnras, 358, 211 

\bibitem[Ruhlen et al.(2011)]{ruhlen2011} Ruhlen, L., Smith, D.~M., \& Swank, J.~H.\ 2011, \apj, 742, 75 

\bibitem[Serlemitsos et al. (2007)]{serlemitsos2007} Serlemitsos, P.~J., Soong, Y., Chan, K.-W., et al.\ 2007, \pasj, 59, 9 

\bibitem[Shakura \& Sunyaev (1973)]{shakura1973} Shakura, N.I., \& Sunyaev, R. A. 1973, \aap, 24, 337

\bibitem[Shimura \& Takahara(1995)]{shimura1995} Shimura, T., \& Takahara, F.\ 1995, \apj, 445, 780 

\bibitem[Steiner et al. (2009)]{steiner2009} Steiner, J. F., Narayan, R., McClintock, J. E., \& Ebisawa, K., 2009, \pasp, 121, 1279

\bibitem[Steiner et al.(2010)]{steiner2010} Steiner, J.~F., McClintock, J.~E., Remillard, R.~A., et al.\ 2010, \apjl, 718, L117 

\bibitem[Steiner et al. (2012)]{steiner2012} Steiner, J.~F., Reis, R.~C., Fabian, A.~C., et al.\ 2012, \mnras, 427, 2552 

\bibitem[Takahashi et al. (2007)]{takahashi2007} Takahashi, T., Abe, K., Endo, M., et al.\ 2007, \pasj, 59, 35 

\bibitem[Takahashi et al.(2008)]{takahashi2008} Takahashi, H., Fukazawa, Y., Mizuno, T., et al.\ 2008, \pasj, 60, 69 

\bibitem[Tamura et al.(2012)]{tamura2012} Tamura, M., Kubota, A., Yamada, S., et al.\ 2012, \apj, 753, 65 

\bibitem[Wilms et al.(2000)]{wilms2000} Wilms, J., Allen, A., \& McCray, R.\ 2000, \apj, 542, 914 

\bibitem[Yamada et al. (2012)]{yamada2012} Yamada, S., et al. 2012, \pasj, 64, 53 

\bibitem[Zdziarski et al.(1993)]{zdziarski1993} Zdziarski, A.~A., Lightman, A.~P., \& Maciolek-Niedzwiecki, A.\ 1993, \apjl, 414, L93 

\bibitem[Zdziarski et al.(1996)]{zdziarski1996} Zdziarski, A.~A., Johnson, W.~N., \& Magdziarz, P.\ 1996, \mnras, 283, 193 

\bibitem[Zdziarski \& Gierli{\'n}ski(2004)]{zdziarski2004} Zdziarski, A.~A., \& Gierli{\'n}ski, M.\ 2004, Progress of Theoretical Physics Supplement, 155, 99 

\end{thebibliography}
\end{document}